\documentclass[12pt]{iopart}

\def\be{\begin{equation}} 
\def\ee{\end{equation}} 
\def\bea{\begin{eqnarray}} 
\def\eea{\end{eqnarray}} 

\input epsf.tex
\usepackage[dvips]{graphicx} 

\begin{document}

\title{String Gas Cosmology: Progress and Problems}

\author{Robert H. Brandenberger}

\address{Physics Department, McGill University, Montreal, QC, H3A 2T8, Canada}
\ead{rhb@physics.mcgill.ca}
\begin{abstract}

String Gas Cosmology is a model of the evolution of the very early universe
based on fundamental principles and key new degrees of freedom of string theory
which are different from those of point particle field theories. In String Gas Cosmology
the universe starts in a quasi-static Hagedorn phase during which space is
filled with a gas of highly excited string states. Thermal fluctuations of this
string gas lead to an almost scale-invariant spectrum of curvature fluctuations.
Thus, String Gas Cosmology is an alternative to cosmological inflation as a theory
for the origin of structure in the universe. This short review focuses on the building
blocks of the model, the predictions for late time cosmology, and the
main problems which the model faces.
\end{abstract}

\submitto{\CQG}

\maketitle

\section{Introduction}

Standard Big Bang (SBB) cosmology teaches us that the further we go back
in time in the history of the universe, the higher is the energy scale of the physics
which is responsible for the detailed dynamical evolution of the early universe.
Whereas at the present time, matter can be described by a collection of perfect
fluids, already at the time of recombination atomic physics processes are
crucial. Even earlier on, nuclear physics effects become important and allow
the SBB to explain the origin and abundances of the light elements. Particle
physics processes are crucial at even earlier times and allow us to construct
models which can explain the observed asymmetry between matter and
antimatter. 

Neither atomic physics, nuclear physics or particle physics effects can change
the conclusion that, as long as space-time is described in terms of Einstein's
theory of General Relativity, the temperature and density increase to
infinite values in finite time as we go back in time. This is the ``Singularity Problem" 
of SBB cosmology. However, at sufficiently high densities it is certainly not
justified to neglect quantum gravity effects. If string theory is indeed the correct
theory of quantum gravity, then string theory must be essential to describe
the universe at the highest densities.

The origin of structure in the universe is a mystery which cannot be
explained in SBB cosmology. Given the wealth of current data on the
non-random distribution of galaxies on cosmological scales and on the
observed anisotropies in the temperature maps of the Cosmic Microwave
Background (CMB), a crucial goal of cosmology has become to come up
with a causal explanation of this data. As realized already a decade
before the invention of Inflationary cosmology \cite{SZ,Peebles},
an approximately scale-invariant spectrum of primordial curvature
fluctuations on scales which at early times (e.g. at the time of recombination)
are super-Hubble is required to explain the observed distribution of
matter on cosmological scales. These same authors realized that any
model which provides such a spectrum of adiabatic fluctuations would
predict angular CMB anisotropies with characteristic acoustic oscillations
on angular scales smaller than about one degree. 

Inflationary cosmology was proposed in 1980 \cite{Guth} (see also
\cite{pre-infl} for related work) as a possible explanation for the large
size, the large entropy, the approximate isotropy and the spatial
flatness of the universe. It was then almost immediately realized \cite{Mukh} 
(see also \cite{pre-flucts} for related work) that inflation can provide
an explanation for the origin of an approximately scale-invariant
spectrum of cosmological fluctuations.

Inflationary cosmology (at least as we mostly understand it today) is
based on coupling the potential energy of a slowly rolling scalar field
to Einstein gravity. Thus, inflation does not eliminate the singularity
problem of SBB cosmology \cite{Borde}. Thus, we still need to address
the question of what happened at times before the hypothetical inflationary
phase. This is where string theory will play a key role.

Inflationary cosmology is faces with other conceptual problems (see
e.g. \cite{RHBrev0, RHBrecent} for in depth discussions of some of
these problems). For example, the wavelength of fluctuations which
we observe today is predicted to be smaller than the Planck length
at the beginning of the inflationary phase in all inflation models in
which the period of inflation is somewhat longer than the minimal
length required for inflation to solve the problems of SBB cosmology
which it is designed to solve. This is the ``trans-Planckian" problem
for fluctuations in inflationary cosmology \cite{RHBrev0, Jerome}.
Furthermore, in simple models of inflation the energy scale at which
inflation takes place is too close to the string scale and Planck scale
to comfortably neglect quantum gravity effects. Finally, the mechanism
to obtain inflationary expansion of space is subject to our ignorance
on how the cosmological constant problem is solves (the driving
force of inflation can be viewed as a temporary cosmological
constant). In light of these problems it is useful to consider the
possibility of an alternative to inflation emerging from (for example)
string cosmology.

It is certainly possible that a string theoretical understanding of the
very early universe will lead to a convincing realization of inflationary
cosmology (see e.g. \cite{stringinflationrevs} for reviews of work along
these lines). Most of the work on string inflation is, however, based on
particle physics quantum field models motivated by string theory rather
than on string theory itself, and thus may be missing crucial aspects
of early universe cosmology which arise due to the specific stringy
nature of matter. 

String Gas Cosmology (SGC) \cite{BV} (see also \cite{Perlt} for
related work and \cite{SGCrevs} for reviews) is a model of early universe 
cosmology based on making use of fundamental principles which
distinguish string theory from point particle theory: the existence of
new stringy degrees of freedom and the resulting new symmetries.
As realized in \cite{BV}, SGC provides the framework for constructing
a nonsingular cosmological model. As realized more recently \cite{NBV, BNPV2},
SGC leads to a mechanism for producing an approximately scale-invariant
spectrum of cosmological fluctuations, and makes the key prediction
\cite{BNPV1} that the spectrum of gravitational wave should have a slight
blue tilt. Thus, SGC emerges as a possible alternative to the inflationary
scenario as an explanation for the observed large-scale structure of the
universe.

In the following, I briefly review SGC. In the following section I present
the basics of SGC. Section 3 explains how SGC leads to late time
curvature fluctuations. The final section highlights challenges which
SGC faces. 
     
\section{Basics of String Gas Cosmology}

String Gas Cosmology is based on coupling a classical background
which includes the graviton and the dilaton fields to a gas of
closed strings (and possibly other basic degrees of freedom of
string theory such as ``branes" \cite{ABE}). All dimensions of space are taken
to be compact, for reasons which will become clear later.
For simplicity, we take all spatial directions to be toroidal and
denote the radius of the torus by $R$. For SGC to be effective in
the way it is currently formulated, there need to be stable or at least
long lived winding modes about all of the spatial dimensions which
are not large today. For the SGC structure formation mechanism of
\cite{NBV} to work, there also need to be stable winding modes
about our three large spatial dimensions. These provide constraints
on the topology of space.
 
Strings have three types of states: {\it momentum modes} which 
represent the center of mass motion of the string, {\it oscillatory modes} which
represent the fluctuations of the strings, and {\it winding
modes} counting the number of times a string wraps the torus.
Since the number of string oscillatory states increases exponentially
with energy, there is a limiting  temperature for a gas of strings in
thermal equilibrium, the {\it Hagedorn temperature} \cite{Hagedorn}
$T_H$. Thus, if we take a box of strings and adiabatically decrease the box
size, the temperature $T$ will never diverge. The fact that $T$ remains
finite as $R$ ranges from $0$ to $\infty$ indicates that the cosmological
singularity can be resolved in SGC.

The second key feature of string theory upon which SGC
is based is {\it T-duality}. Consider the
radius dependence of the energy of the basic string states:
The energy of an oscillatory mode is independent of $R$, momentum
mode energies are quantized in units of $1/R$, i.e.
\be
E_n \, = \, n \mu \frac{{l_s}^2}{R} \, ,
\ee
where $l_s$ is the string length and $\mu$ is the mass per unit length of
a string, but winding mode energies are 
quantized in units of $R$, i.e.
\be
E_m \, = \, m \mu R \, ,
\ee
where both $n$ and $m$ are integers. Thus, under the change
\be
R \, \rightarrow \, 1/R
\ee
in the radius of the torus (in units of  $l_s$)
the energy spectrum of string states is
invariant if winding
and momentum quantum numbers are interchanged
\be
(n, m) \, \rightarrow \, (m, n) \, .
\ee

This symmetry is part of a larger group, the T-duality symmetry group.
The string vertex operators are consistent with this symmetry, and
thus T-duality is a symmetry of perturbative string theory. Postulating
that T-duality extends to non-perturbative string theory leads
\cite{Pol} to the need of adding D-branes to the list of fundamental
objects in string theory. With this addition, T-duality is expected
to be a symmetry of non-perturbative string theory.
Specifically, T-duality will take a spectrum of stable Type IIA branes
and map it into a corresponding spectrum of stable Type IIB branes
with identical masses \cite{Boehm}. 

As discussed in \cite{BV}, the above T-duality symmetry leads to
an equivalence between small and large spaces. This reinforces
the conclusion that there is no cosmological singularity in SGC.
As $R$ decreases, the physically measured size of the universe
will initially decrease but then (once $R$ has passed below the
string length) increase again. The point $R = 0$ will not be
reached in finite observer time.

\begin{figure} 
\includegraphics[height=6cm]{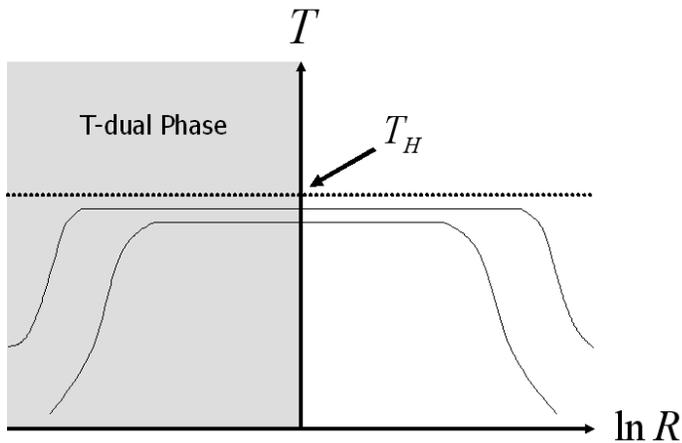}
\caption{The temperature (vertical axis) as a function of
radius (horizontal axis) of a gas of closed strings in thermal
equilibrium. Note the absence of a temperature singularity. The
range of values of $R$ for which the temperature is close to
the Hagedorn temperature $T_H$ depends on the total entropy
of the universe. The upper of the two curves corresponds to
a universe with larger entropy.}
\label{jirofig1}
\end{figure}

Let us consider this evolution in a bit more detail and focus on the
state of string gas matter as $R$ changes. As $R$ decreases from an
initially very large value - maintaining thermal
equilibrium - , the temperature first rises as in
standard cosmology since the string states which are occupied
(the momentum modes) get heavier. However, as the temperature
approaches the Hagedorn temperature, the energy begins to
flow into the oscillatory modes and the increase in temperature
levels off. As the radius $R$ decreases below the string scale,
the temperature begins to decrease as the energy begins to
flow into the winding modes whose energy decreases as $R$
decreases (see Figure \ref{jirofig1}).

The equations that govern the background of string gas cosmology
are not known. The Einstein equations are not the correct
equations since they do not obey the T-duality symmetry of
string theory. Many early studies of string gas cosmology were
based on using the dilaton gravity equations \cite{TV,Ven,Tseytlin},
However, these equations are not satisfactory, either, as will be 
discussed in Section 4.  

\begin{figure} 
  \includegraphics[height=6cm]{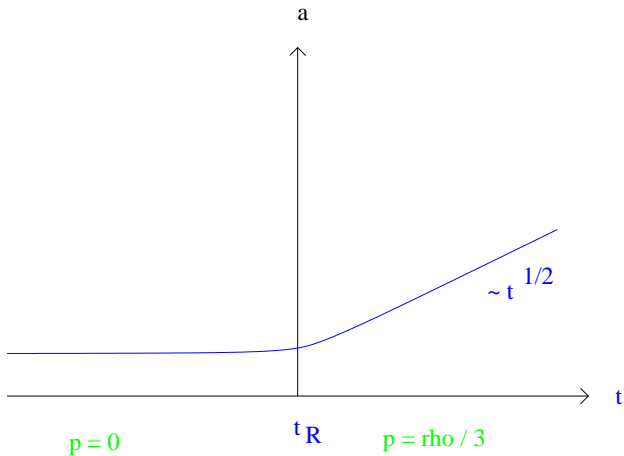}
\caption{The dynamics of string gas cosmology. The vertical axis
represents the scale factor of the universe, the horizontal axis is
time. Along the horizontal axis, the approximate equation of state
is also indicated. During the Hagedorn phase the pressure is negligible
due to the cancellation between the positive pressure of the momentum
modes and the negative pressure of the winding modes, after time $t_R$
the equation of state is that of a radiation-dominated universe.}
\label{timeevol2}
\end{figure}

Some conclusions about the time-temperature relation in string
gas cosmology can be derived based on thermodynamical
considerations alone. One possibility is that  $R$ starts out
much smaller than the self-dual value and increases monotonically.
From Figure \ref{jirofig1} it then follows that the time-temperature curve
will correspond to that of a bouncing cosmology. Alternatively,
it is possible that the universe starts out in a meta-stable state
near the Hagedorn temperature, the {\it Hagedorn phase}, and
then smoothly evolves into an expanding phase dominated by
radiation like in standard cosmology (Figure \ref{timeevol2}). 
SGC is based on this assumption.
Note that we are assuming that not only the scale factor but
also the dilaton is constant in time.  

The transition between the quasi-static Hagedorn phase and the
radiation phase at the time $t_R$ is a consequence of 
the annihilation of string winding modes into string loops. 
Since this process corresponds to the production of radiation, we denote
this time by the same symbol $t_R$ as the time of reheating in inflationary
cosmology. As pointed out in \cite{BV}, this annihilation process
only is possible in at most three large spatial dimensions. This is
a simple dimension counting argument: string world sheets have
measure zero intersection probability in more than four large 
space-time dimensions. Hence, string gas cosmology
may provide a natural mechanism for explaining why there are
exactly three large spatial dimensions. 

One of the important features of SGC is that there is a natural
mechanism to stabilize the size and shape moduli of the extra
spatial dimensions, at least in heterotic string theory. The early
Hagedorn phase of SGC is characterized by a gas in which all 
string states are excited. This includes modes with both winding
and momentum numbers about the extra spatial dimensions.
In heterotic string theory the lowest energy state are ``enhanced
symmetry states" which have zero mass when the radius of the
extra spatial section equals the string scale, but non-vanishing winding
and momenta. Following initial work of \cite{Watson1} and
\cite{Watson2} (see also \cite{Eva}), size moduli stabilization was
studied in detail in \cite{Subodh1, Subodh2}, and shape moduli
stabilization in \cite{Edna}. The only modulus which is not stabilized
using the basic ingredients of SGC is the axion-dilaton modulus.
As studied in \cite{Frey}, this modulus can be stabilized by making
use of gaugino condensation, which in turn also leads to supersymmetry
breaking \cite{Sasmita} which is consistent with the current
cosmological constraints. The gaugino condensation mechanism
of dilaton stabilization does not disrupt the natural stabilization of
the size and shape moduli via string gas effects \cite{Frey}.

The size modulus stabilization mechanism \cite{Watson1,Watson2,Subodh1}
is very geometrical: states with non-vanishing momenta and windings
about the extra dimensions lead to a force resisting both contraction (because
then the momenta lead to a large mass) and expansion (because then
the winding states get very heavy), thus yielding a preferential value
of the radius which is given by the string scale (larger if there is a chemical
potential for winding number). In the case of heterotic string theory,
the lightest states for a value of the size modulus close to the string
scale are the enhances symmetry states which contain both windings
and momenta and whose energy at a value of the radion equal to the
string length is given by the state's kinetic energy in the large
dimensions. As studied carefully in \cite{Subodh2}, these states
fix the radion in a way which is consistent with cosmological constraints.
The extra states act as radiation from our four-dimensional space-time
point of view. With very little tuning of the initial density of the
enhanced symmetry states, the resulting contribution of the
states to the current radiation density is below the nucleosynthesis
constraints, and at the same time the mass of the radion fluctuations
about the ground state is above the current lower bound.

\section{Cosmological Fluctuations from String Gas Cosmology}

\begin{figure} 
\includegraphics[height=9cm]{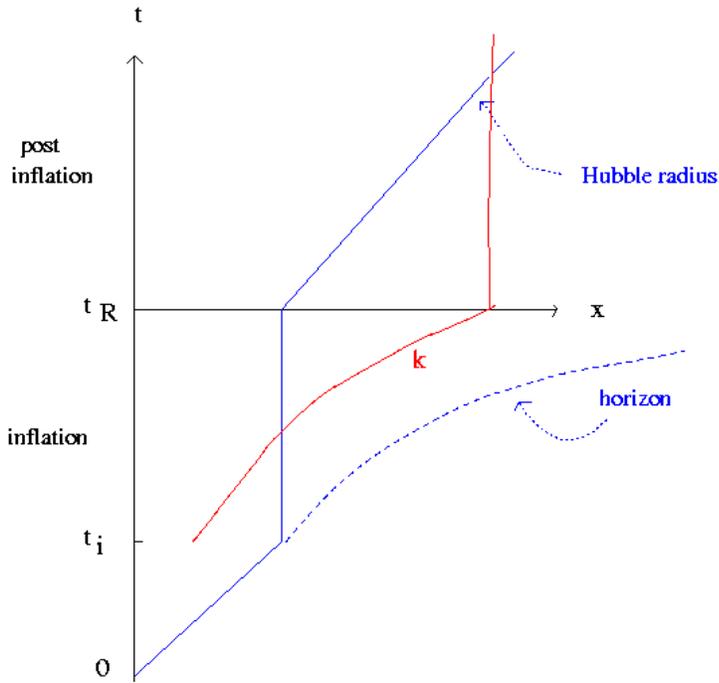}
\caption{Space-time sketch of inflationary cosmology.
The vertical axis is time, the horizontal axis corresponds
to physical distance. The solid line labelled $k$ is the
physical length of a fixed comoving fluctuation scale. The
role of the Hubble radius and the horizon are discussed
in the text.}
\label{infl1}
\end{figure}

In this section we show how thermal fluctuations in the Hagedorn
phase of SGC lead to an almost scale-invariant spectrum of
cosmological perturbations.
It is useful to first remind the reader of the mechanism
by which inflationary cosmology leads to a causal generation 
mechanism for cosmological fluctuations yielding  
an almost scale-invariant spectrum of perturbations.
The space-time diagram of inflationary cosmology is sketched
in Figure \ref{infl1}. 

During the period of inflation, the Hubble radius
$ l_H(t) \, \equiv \, \frac{a}{{\dot a}} $
is approximately constant. In contrast, the physical length
of a fixed co-moving scale (labelled by $k$ in the figure)
is expanding exponentially.
In this way, in inflationary cosmology scales which have
microscopic sub-Hubble wavelengths at the beginning of
inflation are red-shifted to become super-Hubble-scale
fluctuations at the end of the period of inflation.
In the post-inflationary phase of Standard cosmology
the Hubble radius increases linearly in
time, i.e. faster than the physical wavelength corresponding
to a fixed co-moving scale. Thus, scales re-enter the
Hubble radius at late times.

Since inflation red-shifts any classical fluctuations which might
have been present at the beginning of the inflationary phase,
fluctuations in inflationary cosmology are taken to be generated by
quantum vacuum perturbations \cite{Mukh}. The fluctuations begin
in their quantum vacuum state at the onset of inflation. Once the
wavelength exceeds the Hubble radius, squeezing of the
wave-function of the fluctuations sets in (see \cite{MFB,RHBrev1}
for reviews of the theory of cosmological perturbations).
This squeezing plus the de-coherence of the fluctuations due
to the interaction between short and long wavelength modes
generated by the intrinsic non-linearities in both the gravitational and
matter sectors of the theory (see \cite{Martineau,Starob3} for
recent discussions of this aspect and references to previous work)
lead to the classicalization of the fluctuations on super-Hubble
scales.

\begin{figure}  
 \includegraphics[height=.5\textheight]{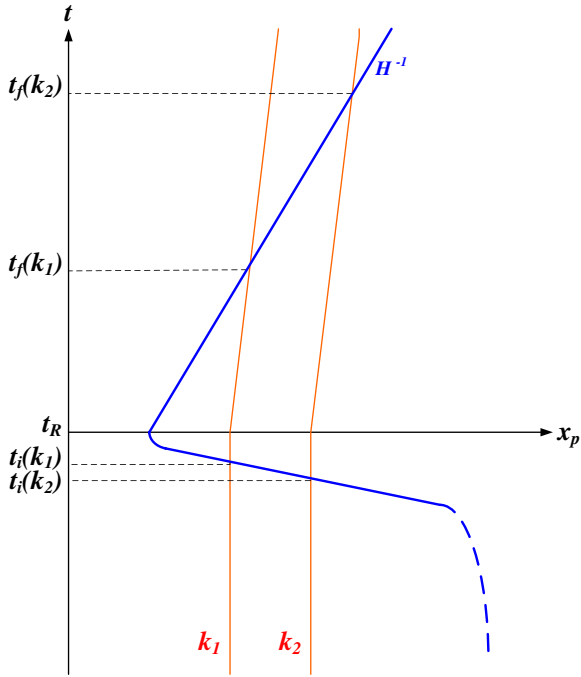}
\caption{Space-time diagram (sketch) showing the evolution of fixed 
co-moving scales in string gas cosmology. The vertical axis is time, 
the horizontal axis is physical distance.  
The solid curve represents the Einstein frame Hubble radius 
$H^{-1}$ which shrinks abruptly to a micro-physical scale at $t_R$ and then 
increases linearly in time for $t > t_R$. Fixed co-moving scales (the 
dotted lines labeled by $k_1$ and $k_2$) which are currently probed 
in cosmological observations have wavelengths which are smaller than 
the Hubble radius before $t_R$. They exit the Hubble 
radius at times $t_i(k)$ just prior to $t_R$, and propagate with a 
wavelength larger than the Hubble radius until they reenter the 
Hubble radius at times $t_f(k)$.}
\label{spacetimenew2}
\end{figure}

The process of generation and evolution of cosmological fluctuations
in SGC is very different. Recall that the evolution of the scale factor
in SGC is as represented in Figure \ref{timeevol2} and leads to a 
space-time diagram as is sketched in Figure \ref{spacetimenew2}.
As in Figure \ref{infl1}, the vertical axis is time and 
the horizontal axis denotes the
physical distance. For times $t < t_R$, 
we are in the static Hagedorn phase and the Hubble radius is
infinite. For $t > t_R$, the Einstein frame 
Hubble radius is expanding as in standard cosmology. The time
$t_R$ is when the string winding modes begin to decay into
string loops, and the scale factor starts to increase, leading to the
transition to the radiation phase of standard cosmology. 

First of all, we see that in SGC all fluctuations emerge at early
times from sub-Hubble scales, as in inflationary cosmology.
This is the first key requirement for a causal generation mechanism
to be possible. Scales are existing the Hubble radius not because
their wavelength is increasing (as in the case of inflation), but rather
because the Hubble radius is decreasing during the transition
between the Hagedorn phase and the expanding radiation phase.
Since the physical wavelength of fluctuations is constant in the
Hagedorn phase, the Trans-Planckian problem for fluctuations
does not arise: if we follow scale which corresponds to the current Hubble
radius back to the time $t_R$, then - taking the Hagedorn
temperature to be of the order $10^{16}$ GeV - we obtain
a length of about 1 mm. Compared to the string scale and the
Planck scale, this is in the far infrared.

The fact that the Hagedorn phase is static and dominated
by a gas of strings lead to a generation mechanism for
fluctuations which is very different from that in inflationary cosmology,
Instead of quantum vacuum perturbations we have thermal
fluctuations, and thermal fluctuations not of a particle gas, but
of a gas of strings. It was realized in \cite{NBV} that this mechanism
yields approximately scale-invariant curvature perturbations
at late times.

The main steps in the computation of cosmological fluctuations
in SGC is as follows
\cite{NBV} (see \cite{BNPV2} for a more detailed description).
For a fixed co-moving scale with wavenumber $k$ we compute the matter
fluctuations while the scale is sub-Hubble (and therefore gravitational
effects are sub-dominant). When the scale exits the Hubble radius
at time $t_i(k)$ we use the gravitational constraint equations to
determine the induced metric fluctuations, which are then propagated
to late times using the usual equations of gravitational perturbation
theory. Since the scales we are interested
in are in the far infrared, we use the Einstein constraint equations for
fluctuations.

We write the metric including cosmological perturbations
(scalar metric fluctuations) and gravitational waves in the
following form (using conformal time $\eta$ which is related
to the physical time $t$ via $dt = a(t) d\eta$) 
\be \label{pertmetric}
d s^2 \, = \, a^2(\eta) \left\{(1 + 2 \Phi)d\eta^2 - [(1 - 
2 \Phi)\delta_{ij} + h_{ij}]d x^i d x^j\right\} \, . 
\ee 
Here, we have adopted the longitudinal gauge and we have taken 
matter to be free of anisotropic stress. The spatial
tensor $h_{ij}({\bf x}, t)$ is transverse and traceless and represents 
gravitational waves. 

Note that in contrast to the case of slow-roll inflation, scalar metric
fluctuations and gravitational waves are generated by matter
at the same order in cosmological perturbation theory. This could
lead to the expectation that the amplitude of gravitational waves
in string gas cosmology should be generically larger than in inflationary
cosmology. This expectation, however, is not realized \cite{BNPV1}
since there is a different mechanism which suppresses the gravitational
waves relative to the density perturbations (namely the fact
that the gravitational wave amplitude is set by the amplitude of
the pressure, and the pressure is suppressed relative to the
energy density in the Hagedorn phase).

Assuming that the fluctuations are described by the perturbed Einstein
equations (they are {\it not} if the dilaton is not fixed 
\cite{Betal,KKLM}), then the spectra of cosmological perturbations
$\Phi$ and gravitational waves $h$ are given by the energy-momentum 
fluctuations in the following way \cite{BNPV2}
\be \label{scalarexp} 
\langle|\Phi(k)|^2\rangle \, = \, 16 \pi^2 G^2 
k^{-4} \langle\delta T^0{}_0(k) \delta T^0{}_0(k)\rangle \, , 
\ee 
where the pointed brackets indicate expectation values, and 
\be 
\label{tensorexp} \langle|h(k)|^2\rangle \, = \, 16 \pi^2 G^2 
k^{-4} \langle\delta T^i{}_j(k) \delta T^i{}_j(k)\rangle \,, 
\ee 
where the right hand side of (\ref{tensorexp}) is the 
average over the correlation functions with $i \neq j$, and
$h$ is the amplitude of the gravitational waves \footnote{The
gravitational wave tensor $h_{i j}$ can be written as the
amplitude $h$ multiplied by a constant polarization tensor.}.
 
To determine the spectrum of
scalar metric fluctuations, we first calculate the 
root mean square energy density fluctuations in a sphere of
radius $R = k^{-1}$. For a system in thermal equilibrium they 
are given by the specific heat capacity $C_V$ via 
\be \label{cor1b}
\langle \delta\rho^2 \rangle \,  = \,  \frac{T^2}{R^6} C_V \, . 
\ee 
The specific  heat of a gas of closed strings
on a torus of radius $R$ can be derived from the partition
function of a gas of closed strings. This computation was
carried out in \cite{Deo} and yields
\be \label{specheat2b} 
C_V  \, \approx \, 2 \frac{R^2/\ell^3}{T \left(1 - T/T_H\right)}\, . 
\ee 
The specific heat capacity scales holographically with the size
of the box. This result follows rigorously from evaluating the
string partition function in the Hagedorn phase. The result, however,
can also be understood heuristically: in the Hagedorn phase the
string winding modes are crucial. These modes look like point
particles in one less spatial dimension. Hence, we expect the
specific heat capacity to scale like in the case of point particles
in one less dimension of space \footnote{We emphasize that it was
important for us to have compact spatial dimensions in order to
obtain the winding modes which are crucial to get the holographic
scaling of the thermodynamic quantities.}.

With these results, the power spectrum $P(k)$ of scalar metric fluctuations can
be evaluated as follows
\bea \label{power2} 
P_{\Phi}(k) \, & \equiv & \, {1 \over {2 \pi^2}} k^3 |\Phi(k)|^2 \\
&=& \, 8 G^2 k^{-1} <|\delta \rho(k)|^2> \, . \nonumber \\
&=& \, 8 G^2 k^2 <(\delta M)^2>_R \nonumber \\ 
               &=& \, 8 G^2 k^{-4} <(\delta \rho)^2>_R \nonumber \\
&=& \, 8 G^2 {T \over {\ell_s^3}} {1 \over {1 - T/T_H}} 
\, , \nonumber 
\eea 
where in the first step we have used (\ref{scalarexp}) to replace the 
expectation value of $|\Phi(k)|^2$ in terms of the correlation function 
of the energy density, and in the second step we have made the 
transition to position space 

The first conclusion from the result (\ref{power2}) is that the spectrum
is approximately scale-invariant ($P(k)$ is independent of $k$). It is
the `holographic' scaling $C_V(R) \sim R^2$ which is responsible for the
overall scale-invariance of the spectrum of cosmological perturbations.
However, there is a small $k$ dependence which comes from the fact
that in the above equation for a scale $k$ 
the temperature $T$ is to be evaluated at the
time $t_i(k)$. Thus, the factor $(1 - T/T_H)$ in the 
denominator is responsible 
for giving the spectrum a slight dependence on $k$. Since
the temperature slightly decreases as time increases at the
end of the Hagedorn phase, shorter wavelengths for which
$t_i(k)$ occurs later obtain a smaller amplitude. Thus, the
spectrum has a slight red tilt.

Now let us turn to the key prediction with which SGC can be
differentiated observationally from inflation. It concern the
tilt in the spectrum of gravitational waves.
As discovered in \cite{BNPV1}, the spectrum of gravitational
waves is also nearly scale invariant. However, in the expression
for the spectrum of gravitational waves the factor $(1 - T/T_H)$
appears in the numerator, thus leading to a slight blue tilt in
the spectrum. This contrasts with the predictions of inflationary
models, where quite generically a slight red
tilt for gravitational waves results. The physical reason for the blue
tilt of the spectrum of gravitational waves in SGC is that
large scales exit the Hubble radius earlier when the pressure
and hence also the off-diagonal spatial components of $T_{\mu \nu}$
are closer to zero.

The method for calculating the spectrum of gravitational waves
follows what was presented above for the scalar metric fluctuations. 
For a mode with fixed co-moving
wavenumber $k$, we compute the correlation function of the
off-diagonal spatial elements of the string gas energy-momentum
tensor at the time $t_i(k)$ when the mode exits the Hubble radius
and use (\ref{tensorexp}) to infer the amplitude of the power
spectrum of gravitational waves at that time. We then
follow the evolution of the gravitational wave power spectrum
on super-Hubble scales for $t > t_i(k)$ using the equations
of general relativistic perturbation theory.

The power spectrum of the tensor modes is given by (\ref{tensorexp}):
\be \label{tpower1}
P_h(k) \, = \, 16 \pi^2 G^2 k^{-4} k^3
\langle\delta T^i{}_j(k) \delta T^i{}_j(k)\rangle
\ee
for $i \neq j$. Note that the $k^3$ factor multiplying the momentum
space correlation function of $T^i{}_j$ gives the position space
correlation function $C^i{}_j{}^i{}_j(R)$ , namely the root mean 
square fluctuation of $T^i{}_j$ in a region of radius $R = k^{-1}$ 
(the reader who is skeptical about this point is invited to check 
that the dimensions work out properly). Thus,
\be \label{tpower2}
P_h(k) \, = \, 16 \pi^2 G^2 k^{-4} C^i{}_j{}^i{}_j(R) \, .
\ee
The correlation function $C^i{}_j{}^i{}_j$ on the right hand side
of the above equation follows from the thermal closed string
partition function and was computed explicitly in
\cite{Ali} (see also \cite{BNPV2}). We obtain
\be \label{tpower3}
P_h(k) \, \sim \, 16 \pi^2 G^2 {T \over {l_s^3}}
(1 - T/T_H) \ln^2{\left[\frac{1}{l_s^2 k^2}(1 -
T/T_H)\right]}\, ,
\ee
which, for temperatures close to the Hagedorn value reduces to
\be \label{tresult}
P_h(k) \, \sim \,
\left(\frac{l_{Pl}}{l_s}\right)^4 (1 -
T/T_H)\ln^2{\left[\frac{1}{l_s^2 k^2}(1 - T/T_H)\right]} \, .
\ee
This shows that the spectrum of tensor modes is - to a first
approximation, namely neglecting the logarithmic factor and
neglecting the $k$-dependence of $T(t_i(k))$ - scale-invariant. 

On super-Hubble scales, the amplitude $h$ of the gravitational waves
is constant. The wave oscillations freeze out when the wavelength
of the mode crosses the Hubble radius. As in the case of scalar metric
fluctuations, the waves are squeezed. Whereas the wave amplitude remains
constant, its time derivative decreases. Another way to see this
squeezing is to change variables to 
\be
\psi(\eta, {\bf x}) \, = \, a(\eta) h(\eta, {\bf x})
\ee
in terms of which the action has canonical kinetic term. The action
in terms of $\psi$ becomes
\be
S \, = \, {1 \over 2} \int d^4x \left( {\psi^{\prime}}^2 -
\psi_{,i}\psi_{,i} + {{a^{\prime \prime}} \over a} \psi^2 \right)
\ee
from which it immediately follows that on super-Hubble scales
$\psi \sim a$. This is the squeezing of gravitational 
waves \cite{Grishchuk}.
Since there is no $k$-dependence in the squeezing factor, the
scale-invariance of the spectrum at the end of the Hagedorn phase
will lead to a scale-invariance of the spectrum at late times.

Note that in the case of string gas cosmology, the squeezing
factor $z(\eta)$ does not differ substantially from the
squeezing factor $a(\eta)$ for gravitational waves. In the
case of inflationary cosmology, $z(\eta)$ and $a(\eta)$
differ greatly during reheating, leading to a much larger
squeezing for scalar metric fluctuations, and hence to a
suppressed tensor to scalar ratio of fluctuations. In the
case of string gas cosmology, there is no difference in
squeezing between the scalar and the tensor modes. 

Let us return to the discussion of the spectrum of gravitational
waves. The result for the power spectrum is given in
(\ref{tresult}), and we mentioned that to a first approximation this
corresponds to a scale-invariant spectrum. As realized in
\cite{BNPV1}, the logarithmic term and the $k$-dependence of
$T(t_i(k))$ both lead to a small blue-tilt of the spectrum. This
feature is characteristic of our scenario and cannot be reproduced
in inflationary models. In inflationary models, the amplitude of
the gravitational waves is set by the Hubble constant $H$. The
Hubble constant cannot increase during inflation, and hence no
blue tilt of the gravitational wave spectrum is possible.

A heuristic way of understanding the origin of the slight blue tilt
in the spectrum of tensor modes
is as follows. The closer we get to the Hagedorn temperature, the
more the thermal bath is dominated by long string states, and thus
the smaller the pressure will be compared to the pressure of a pure
radiation bath. Since the pressure terms (strictly speaking the
anisotropic pressure terms) in the energy-momentum tensor are
responsible for the tensor modes, we conclude that the smaller the
value of the wavenumber $k$ (and thus the higher the temperature
$T(t_i(k))$ when the mode exits the Hubble radius, the lower the
amplitude of the tensor modes. In contrast, the scalar modes are
determined by the energy density, which increases at $T(t_i(k))$ as
$k$ decreases, leading to a slight red tilt.

The reader may ask about the predictions of string gas cosmology
for non-Gaussianities. The answer is \cite{SGNG} that the
non-Gaussianities from the thermal string gas perturbations
are Poisson-suppressed on scales larger than the thermal
wavelength in the Hagedorn phase. However, if the spatial
sections are initially large, then it is possible that a network
of cosmic superstrings \cite{Witten} will be left behind. These
strings - if stable - would achieve a scaling solution (constant
number of strings crossing each Hubble volume at each
time). Such strings give rise
to linear discontinuities in the CMB temperature maps \cite{KS},
lines which can be searched for using edge detection
algorithms such as the Canny algorithm (see \cite{Amsel}
for recent feasibility studies).

\section{Challenges for String Gas Cosmology} 

As as been argued in the above sections, SGC is an interesting
model of early universe cosmology making use of fundamental
principles of string theory which are not used in the standard
string-motivated field theory approaches to string cosmology (most
of string inflation model building falls into that category).
SGC realized the hope that string theory will lead to a non-singular
cosmology since there is a duality between large and small spatial
sections \cite{BV}. SGC provides a natural solution to the stabilization
of size and shape moduli \cite{Subodh2}: these moduli are stabilized
without the need to add any extra ingredients to the model.
One extra ingredient is required in order to stabilize the dilaton -
gaugino condensation is one possibility \cite{Frey}.

Thermal fluctuations of the string gas during an early quasi-static
Hagedorn phase lead to an almost scale-invariant power spectrum
of cosmological perturbations. Thus, SGC is an alternative to
cosmological inflation for explaining the origin of structure in the
universe. The key prediction of SGC is a slight blue tilt in the
spectrum of gravitational waves, whereas inflation generically
produces a slight red tilt. 

The main problem of the current implementation of SGC
is that it does not provide a quantitative model for the Hagedorn phase.
This phase must be quasi-static (including a stable dilaton) and
last for a sufficient length of time to be able to set up thermal
equilibrium on scales which today become comparable to the
current Hubble radius (if the string scale is close to the
scale of Grand Unification), this length is of the order of $1 {\rm{mm}}$).
The criticisms of string gas cosmology raised in \cite{KKLM,Kaloper} are
based on assuming that the background is described by dilaton
gravity.

The problem is that we currently have no framework for describing
the Hagedorn phase mathematically. The Einstein action is
obviously inapplicable since it conflicts with the T-duality
symmetry which is crucial to string theory. Neither can  dilaton
gravity provide suitable background equations. Firstly, as
the dilaton changes from weak to strong coupling, the nature
of the light states in the string spectrum changes and it becomes
inconsistent to model matter as a gas of strings. 
More importantly, at high densities such as the Hagedorn density 
the Einstein term in the gravitational action of both Einstein and
dilaton gravity will no longer be the dominant terms in a
derivative expansion of the action. This is known from all
approaches to quantum gravity. Thus, we cannot expect
that a simple effective action such as that of dilaton gravity will apply. 
Nevertheless, to make the string gas cosmology scenario into a real 
theory, it is crucial to obtain a good understanding of the background 
dynamics. For some initial steps in this direction see \cite{Kanno1}.
Another study of this problem was given in \cite{Sduality}.

A second problem of string gas cosmology is the size problem (and
the related entropy problem). If the string scale is about $10^{17} {\rm GeV}$
as is preferred in early heterotic superstring models, then the radius
of the universe during the Hagedorn phase must be many orders
of magnitude larger than the string scale. Without embedding string
gas cosmology into a bouncing cosmology, it seems unnatural to
demand such a large initial size. This problem disappears if the
Hagedorn phase is preceded by a phase of contraction, as in the
model of \cite{Biswas2}. In this case, however, it is non-trivial to arrange
that the Hagedorn phase lasts sufficiently long to maintain thermal
equilibrium over the required range of scales.

It should be noted, however, that some of the conceptual problems
of inflationary cosmology such as the trans-Planckian problem for
fluctuations, do not arise in string gas cosmology. As in the case
of the matter bounce scenario, the basic mechanism of the scenario
is insensitive to what sets the cosmological constant to its
observed very small value.

\ack

The author is supported in part by an NSERC Discovery Grant, by a Killam Research
Fellowship and by funds from the Canada Research Chair program.

\end{document}